\documentclass[a4paper]{article}
\usepackage{amsmath}
\usepackage{amssymb}
\usepackage{amsthm}
\usepackage{cite}
\usepackage{color}
\usepackage{fullpage}
\usepackage{graphicx}
\usepackage{numprint}
\usepackage{subfig}
\usepackage{hyperref}

\npthousandsep{,}\npthousandthpartsep{}\npdecimalsign{.}

\renewcommand{\P}[0]{P}
\newcommand{\prob}[1]{\P\left(#1\right)}
\newcommand{\F}[0]{\mathcal{F}}
\newcommand{\D}[0]{D}

\newcommand{\expect}[1]{E\left(#1\right)}
\newcommand{\cond}[2]{#1 \, | \, #2}

\newcommand{\N}[0]{\mathbb{N}}
\newcommand{\R}[0]{\mathbb{R}}
\newcommand{\directset}[1]{\left\{ #1 \right\}}
\newcommand{\setr}[2]{\directset{#1 \left| \, #2 \right.}}
\newcommand{\abs}[1]{\left| #1 \right|}
\newcommand{\sign}[1]{\mathrm{sgn} \left( #1 \right)}
\newcommand{\ldef}[0]{:=}
\newcommand{\rdef}[0]{=:}
\newcommand{\num}[1]{\numprint{#1}}
\newcommand{\percent}[1]{\numprint{#1}\%}

\renewcommand{\eqref}[1]{\hyperref[eq:#1]{(\ref*{eq:#1})}}
\newcommand{\tabref}[1]{\hyperref[tab:#1]{Table~\ref*{tab:#1}}}

\newcommand{\mysection}[2]{\section{#2}\label{sec:#1}}
\newcommand{\mysubsection}[2]{\subsection{#2}\label{subsec:#1}}
\newcommand{\secref}[1]{\hyperref[sec:#1]{Section~\ref*{sec:#1}}}
\newcommand{\subsecref}[1]{\hyperref[subsec:#1]{Subsection~\ref*{subsec:#1}}}

\newcommand{\lemmaref}[1]{\hyperref[lemma:#1]{Lemma~\ref*{lemma:#1}}}
\newcommand{\corref}[1]{\hyperref[cor:#1]{Corollary~\ref*{cor:#1}}}
\newcommand{\propref}[1]{\hyperref[prop:#1]{Proposition~\ref*{prop:#1}}}
\newcommand{\thref}[1]{\hyperref[thm:#1]{Theorem~\ref*{thm:#1}}}
\newcommand{\defref}[1]{\hyperref[def:#1]{Definition~\ref*{def:#1}}}

\newtheorem{realLemma}{Lemma}
\newenvironment{lemma}[1]
  {\begin{realLemma}\label{lemma:#1}}
  {\end{realLemma}}
\newtheorem{realCorollary}[realLemma]{Corollary}
\newenvironment{corollary}[1]
  {\begin{realCorollary}\label{cor:#1}}
  {\end{realCorollary}}
\newtheorem{realProp}[realLemma]{Proposition}
\newenvironment{proposition}[1]
  {\begin{realProp}\label{prop:#1}}
  {\end{realProp}}
\newtheorem{realThm}[realLemma]{Theorem}
\newenvironment{theorem}[1]
  {\begin{realThm}\label{thm:#1}}
  {\end{realThm}}
\newtheorem{realDef}[realLemma]{Definition}
\newenvironment{definition}[1]
  {\begin{realDef}\label{def:#1}}
  {\end{realDef}}

\newcommand{\figref}[1]{\hyperref[fig:#1]{Figure~\ref*{fig:#1}}}
\newcommand{\subfigref}[2]{\hyperref[subfig:#1:#2]{Figure~\ref*{subfig:#1:#2}}}

\newcommand{\includefig}[4]
  {
    \begin{figure}
    \begin{center}
    \includegraphics[#2]{#1}
    \end{center}
    \caption{#4}
    \label{fig:#3}
    \end{figure}
  }
\newcommand{\figid}[0]{}
\newcommand{\includesubfig}[3]
  {
    \begin{figure}
    \renewcommand{\figid}[0]{#1}
    \begin{center}
      #3
    \end{center}
    \caption{#2}
    \label{fig:#1}
    \end{figure}
  }
\newcommand{\subfig}[4]
  {
    \subfloat[#4]
      {\label{subfig:\figid:#3}\includegraphics[width=#1\textwidth]{#2}}
  }

\title{Game-Theoretic Randomness for Blockchain Games}

\author{Daniel Kraft \\
XAYA Project \\
Email: daniel@xaya.io \\
ORCID: 0000-0002-0862-5350}
\date{January 18th, 2019}

\begin{document}
\maketitle

\begin{abstract}
In this paper, we consider the problem of generating fair
randomness in a deterministic, multi-agent context (for instance,
a decentralised game built on a blockchain).
The existing state-of-the-art approaches are either susceptible to
manipulation if the stakes are high enough, or they are not generally
applicable (specifically for massive game worlds as opposed to
games between a small set of players).
We propose a novel method based on game theory:  By allowing
agents to bet on the outcomes of random events against the miners
(who are ultimately responsible for the randomness), we are able to align
the incentives so that the distribution of random events is skewed only
slightly even if miners are trying to maximise their profit
and engage in block withholding to cheat in games.

\vspace{1em}
\end{abstract}

\mysection{introduction}{Introduction}

With the creation of Bitcoin \cite{bitcoin} in 2008, its pseudonymous inventor
Satoshi Nakamoto solved a previously impossible problem:  It became possible
for fully decentralised P2P networks to reach consensus about the current
state of a distributed ledger.  This allowed the creation of secure digital
money without any trusted intermediary, and was without doubt a revolutionary
progress of technology.

But Nakamoto consensus is not limited to building cryptocurrencies
like Bitcoin.  The same mechanism can be applied to reach consensus
about other things as well in a decentralised setting.  The first such
application was Namecoin \cite{namecoin}, where Nakamoto consensus is
used to establish who was the first to register a particular name
(e.g. a domain name or pseudonym) in the system
and who is the ``rightful'' owner of it.
Recently, there has also been a lot of interest and activity in the field of
\emph{blockchain games}, where Nakamoto consensus is applied to
an online game.  This idea was pioneered by Huntercoin
\cite{huntercoin} \cite{blockchainGames} in 2014, and became widely
known with the launch of CryptoKitties \cite{cryptoKitties} in 2017.

In this paper, we want to tackle one problem of specific interest
in blockchain games:  How to produce fair and secure random numbers.
By the nature of blockchain systems and the need for a consensus among
network participants, all computations need to be deterministic.
This includes events in games that are meant to be ``random''.  Furthermore,
since blockchain applications often involve monetary stakes, it is important
to produce random numbers in a way that cannot be manipulated or predicated,
giving unfair advantages to certain participants in a game.

There are currently two dominant approaches for the generation of random numbers
in blockchain games:  They can be based off block hashes or a hash-commitment
scheme can be used to generate provably-fair random numbers among a
well-defined set of players (e.g. a casino and a player).
Both methods, however, have certain drawbacks:  The first can be
manipulated by miners if the stake in a game is high enough; the second
is only applicable in some situations and, notably, not to MMO-type
games like Huntercoin.
This will be discussed in more detail in \secref{state of the art}.

In \secref{betting} below, we propose a novel alternative
method, which is based on block hashes and generally
applicable, but uses a specific \emph{betting mechanism} to punish dishonest
miners.  Our game-theoretic
analysis in \secref{analysis} shows that
this does, indeed, align miner incentives with those of players of a
blockchain game.
(For the main result, see \thref{Pd bound}.)
Manipulation of random numbers by miners is
strongly discouraged, so that miners are much more honest and
the game play will be much fairer for everyone.

\mysection{background}{Blockchain Background}

In this section, we want to give a brief overview of the blockchain
background that is necessary for the remainder of this paper.
A more general description of the
basic mechanisms employed by a blockchain using Nakamoto consensus
can be found in the seminal paper \cite{bitcoin} or the more
extensive book \cite{masteringBitcoin}.

\mysubsection{mining}{Proof-of-Work Mining}

A \emph{blockchain} is an append-only data structure, where
new data (contained in \emph{blocks}) is added over time to the end
of an ever-growing list of previous blocks.  If a network participant
(called a \emph{miner}) wants to append a new block to the list, they need to
spend computational resources to produce a \emph{proof-of-work} (PoW); this is
a data structure that is expensive to compute, but where other participants
can cheaply verify that a certain amount of computation was done to create it
(see also \cite{hashcash}).
In the case of Bitcoin (and most other related blockchains based on this
principle), such a PoW is computed by brute-forcing a partial collision of
a cryptographic hash function.

This process, called \emph{mining}, is one of the key ingredients for the
Nakamoto consensus:  By having to spend real-world resources (energy to
power the computation), miners have an economic incentive to behave
``well'' and produce a single chain of blocks that everyone agrees on
rather than working on different versions that compete with each other.

\mysubsection{state transitions}{State Transitions}

The blockchain as data structure and the mining process are ultimately
used to allow the network to \emph{reach consensus about some state}.
In the case of Bitcoin, this state is roughly speaking the current ledger
of bitcoin balances.  (In reality, Bitcoin does not track individual
``balances'' but instead unspent transaction outputs in the so-called
\emph{UTXO set}.  But for the context of this paper, this technical
distinction is not important.)

This is achieved by coupling the current state with the blockchain
through pre-defined rules for \emph{state transitions}:  The blockchain
with its consensus mechanism establishes a well-defined series of transactions
(requested changes to the state) made by the participants.  For each new block
of transactions that is generated, the previous state is updated according to
the state-transition rules based on the transactions in that new block.

It is important to note here that the blockchain itself stores
\emph{the transactions} (i.e. actions by the participants) and
\emph{not the state}.  This is enough, since the state is uniquely defined
already by the series of transactions made and the state-transition rules,
so that it can be computed independently and stored as needed by every
network participant.

The actual state-transition rules used in blockchain networks are quite
diverse.  For ``basic'' blockchains like Bitcoin or Namecoin, they are
relatively simple.
In the case of Huntercoin, the state transition also encodes the
rules of the embedded game world, including harvesting of resources
in the world and basic combat between players.
For Ethereum \cite{ethereum}, state transitions are computed
by executing Turing-complete byte code on the EVM (Ethereum Virtual Machine),
so that the state itself can contain arbitrary programs that determine further
rules (similar to the von-Neumann architecture of modern computers).

It is even possible to use a single blockchain as the underlying storage
layer for data that is then used to compute multiple distinct states
according to different rules.  This is done by overlay protocols like
Mastercoin \cite{mastercoin} (now called Omni Layer) or
Counterparty \cite{counterparty}.
The XAYA blockchain \cite{xayaGames}, which is a generalisation of
the Huntercoin model, is even specifically built to be the data layer for
different sets of state-transition rules.
This is illustrated in \figref{xaya state update}.

\includefig{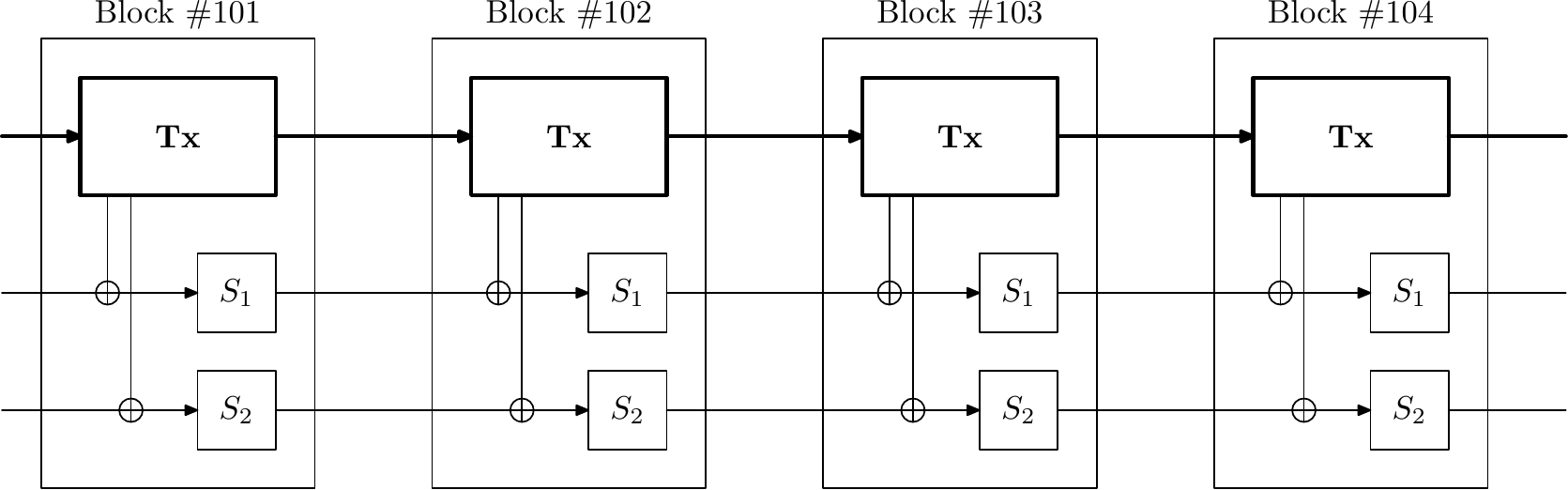}{width=\textwidth}{xaya state update}
  {Illustration of a single blockchain storing transactions (bold outline)
   from which two different states $S_1$ and $S_2$ associated
   to each block are computed according to two different sets of
   transition rules.}

\mysubsection{games}{Blockchain Games}

The network state and its state-transition rules as described above can
now be used specifically to build (multi-player online) \emph{games}.
Using a blockchain for this allows them to be run by the network as a whole,
so that no central game server is needed and instead every participant
computes the current state and verifies that it is correct according to the
game rules.  This has various benefits, including provably-fair game play
and independence from central servers (that can be hacked, go down or simply
be shut off by the company running the game).

Note that the term ``game'' is used in a wide sense for the context of this
paper:  It can refer to MMO-type worlds like Huntercoin, simple trading
games like CryptoKitties, skill games or even gambling in online casinos.
Furthermore, such a game does not even need to be about entertainment.
It can be a purely economic interaction between agents, as long as there
is a predefined and well-known set of rules that govern it.

\mysection{state of the art}{Random Events in Games}

Let us now consider how the state-transition function used in a game
can actually produce \emph{random events}.  Randomness is important for
many types of games, but on the other hand all state transitions in a
blockchain need to be fully deterministic and reproducible for every
participant in the network.

The current state of the art for randomness in blockchain games is
based on two quite different approaches:  Using the hash of the current
block to seed a pseudo-random number generator for the state computation,
or using hash commitments directly between the players to generate
random numbers that are provably fair.
In this section, we will discuss both of these approaches.  It will turn
out that both have nice properties but also drawbacks, and that there
are interesting applications for which neither is fully suited.

\mysubsection{block hash rng}{Block Hashes and Block Withholding}

Since the state transition cannot be \emph{really} random, it can instead
rely on a pseudo-random number generator (PRNG).  If the PRNG is seeded
based on the hash of the current block, then the resulting random numbers
are unpredictable until the block has been mined.
(If the same hash function is used for the PoW algorithm and for
seeding the PRNG, then the seed will have a bias towards low values.
But if that is a problem in a particular situation, it can be easily fixed
by using two different hash functions or, for instance, hashing the
block hash again to compute the seed.)

This approach is straight-forward, and can be applied to introduce randomness
into arbitrary state-transition functions.  Because of that, it is widely used
for blockchain games.  For instance, both Huntercoin (on its own
blockchain) and CryptoKitties (on Ethereum) as well as many other games
and online casinos on the Ethereum platform apply this method.

Unfortunately, this method also has a big defect:  Even though the
outcome of random events is not decided until the block is mined, the
miner who produced the block still is the first who knows the result.
This means that he may decide to simply discard the block instead of
publishing it, particularly when the miner participates in a game as well
and the outcome is disadvantageous for him.
By doing so, the miner obviously has the opportunity cost of losing the
block reward that he would get for the solved block.  But if the stake
in a game is high enough, it may still be worthwhile to withhold the block.
The exact game-theoretic incentive structure for miners that also participate
in blockchain-based casino games has been analysed in \cite{ledgerCasinos}.

For many games, especially if they are small and/or based on a widely
used blockchain like Ethereum, this may be an acceptable risk to take.
But for other applications, the risk of a miner manipulating the randomness
through block withholding can be prohibitive.

\mysubsection{merged mining}{Considerations for Merged Mining}

An alternative to the direct mining process described above in
\subsecref{mining} is \emph{merged mining} \cite{mergedMining}.
Pioneered by Namecoin in 2011, this method allows miners of a parent
blockchain (like Bitcoin) to also mine on a merge-mined blockchain like
Namecoin ``for free''.  That way, it is possible for otherwise smaller
blockchains to gain a big amount of hashing power, giving
a big boost to their security.

The way how this works is as follows:  Instead of computing a PoW
for the current Namecoin block directly, miners construct
a Bitcoin block instead, but include a reference to the Namecoin block in it.
Then, if they find a suitable PoW for the Bitcoin block, this proof also
commits to the Namecoin block.  The Namecoin network is built in such a way
that it accepts this ``indirect'' PoW as well.
This is illustrated in \figref{merged mining}.

\includefig{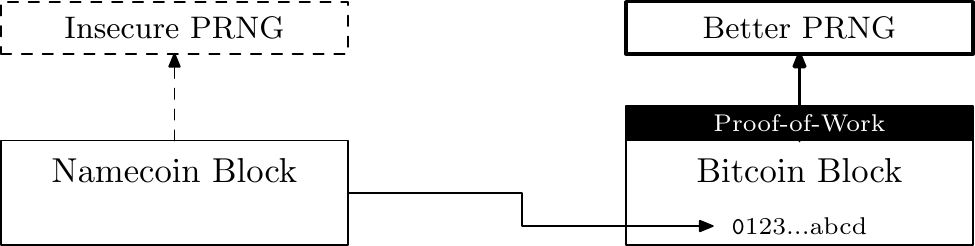}{width=0.8 \textwidth}{merged mining}
  {For a merge-mined blockchain like Namecoin, the PoW is not attached
   to the Namecoin block directly.  Instead, a parent block (Bitcoin) is
   constructed that commits to the Namecoin block's hash, and the PoW is
   computed for the parent block.  Indirectly, it still depends on the
   Namecoin block, and thus can secure it.
   To seed a PRNG in this situation, the parent block (Bitcoin) should
   be used (bold) instead of the Namecoin block itself (dashed).}

Merged mining can be very beneficial for the security of a blockchain.
However, its use mandates a tweak to the use of block hashes to seed PRNGs:
Since the PoW is not directly part of the block itself, the block hash also
does not depend on it.  This means that a miner can cheaply generate multiple
versions of a block and its block hash, before ever starting the
computationally-intensive mining process.  Thus, if the randomness of a game
were based on the block hash, the miner could simply generate a block he
likes first and only then start mining.  The opportunity cost of block
withholding would be completely removed, making it very easy and cheap to
manipulate random numbers.

This problem is straight-forward to fix, though:  Instead of basing random
numbers off the block hash, they have to depend on some quantity that
itself depends on the actual PoW.  So by simply using, for instance, the
block hash \emph{of the parent chain}, the same opportunity cost as with a
non-merge-mined blockchain is restored.
The XAYA blockchain, which can be merge mined with Bitcoin,
includes exactly this fix in its \texttt{rngseed} mechanism.

\mysubsection{hash commitments}{Hash Commitments}

A completely different method for generating provably-fair randomness in a game
is based on \emph{hash commitments}.  For the simple case of a game between
two players (e.g. one of them could be an online casino), the determination
of a random event (e.g. roulette spin) could look like this:

\begin{enumerate}

\item
Both players choose a secret random number, $N_1$ and $N_2$.
The random event's outcome will be based on a combination of
$N_1$ and $N_2$, e.g. on the hash $H(N_1 | N_2)$ of a concatenation of both.

\item
The players share the \emph{hashes} of their secrets with each other,
$H(N_1)$ and $H(N_2)$.

\item
At this point in time, the secrets and with them the outcome of the
random event are fixed.  None of the players can change their value
anymore without the other noticing, but none can predict the the outcome
yet without knowing the other's secret.

\item
After both players have chosen a secret and committed to it, both reveal their
preimages $N_1$ and $N_2$.  Now both can verify that the other followed the
protocol and both can compute and verify $H(N_1 | N_2)$ to determine
the random event independently.

\end{enumerate}

Of course, the full protocol for some game based on this approach will
likely also include steps where the players sign their messages to each other
and/or include them in blockchain transactions, but this is not relevant
for the discussion here.

With a protocol like this, both players are guaranteed provably-fair
random numbers:  As long as the hash function $H$ is cryptographically
secure, none of the players can manipulate or predict the outcome.

Hash-commitment schemes are, of course, not new.  They have been proposed
for cryptographically-secure games long before the invention of Bitcoin,
for instance in \cite{mentalPoker}.
In the Bitcoin ecosystem, SatoshiDice \cite{satoshiDice} was an early gambling
platform that utilised hash commitments to provide users with
provably-fair betting.
Another, more sophisticated example for the use of hash commitments are the
``Fate Channels'' described by FunFair \cite{funFair}.

But while hash-commitment schemes provide provable security against
manipulation, they are unfortunately not applicable in all situations.
They are good for games between a well-defined set of players (e.g. just
a user and a casino), but they cannot be applied in their basic form for
large game worlds with an unspecified set of (currently online) users
like Huntercoin.  Of course, users might be allowed to contribute data
that is integrated into the random-number computation also in these situations.
But since it is not known which users can or want to contribute, this has
to be optional.  Then, however, miners get back the ability to manipulate
the randomness at will, since they will be able to censor reveal
transactions of users which they do not like; ultimately, the miner of a block
will again be the first person to know the outcome, and be able to withhold the
block just as discussed above in \subsecref{block hash rng}.

\mysection{betting}{Betting against Dishonest Miners}

As we have seen before in \secref{state of the art}, none of the existing
approaches to random numbers can provide secure randomness for general
blockchain games.  The usage of block hashes to seed PRNGs works for
all kinds of games, but unfortunately suffers from the risk of block
withholding.
This issue, however, can be fixed by adding an additional feature:
Namely by \textbf{letting users bet ``against'' miners}
as first proposed in \cite{bitcointalkThread}.

Before we describe this idea in more detail, let us take a brief look
at a simple and classical game, rock-paper-scissors.  It is easy to see
(a simple overview can be found in \cite{rockPaperScissors})
that the optimal strategy for
each player there is to pick each choice randomly with probability $1/3$.
As soon as one player deviates from this strategy, she gives an advantage to
her opponent who can then exploit the predictability in her strategy.
Thus, ideally each player should be as unpredictable as possible, which
means randomising the choices as much as possible.

The same can be applied also to randomness based on block hashes:
In particular, consider a game where users of a blockchain can bet
whether some future block hash will be even or odd.  If they bet correctly,
they get some money from the miner who created the block in question
(their bet minus some house edge).
If their bet is wrong, the miner instead wins the amount they wagered.

If a miner produces perfectly randomised block hashes (which is automatically
the case as long as he does not engage in block withholding), then this betting
game adds some variance to their payout, but overall they win due to the
house edge.  But as soon as a miner's blocks have some kind of bias and
are no longer fully random, the betting users can get an advantage by
exploiting this to win money at the cost of the miner.  Thus, the community
at large gets an instrument they can apply to \emph{punish dishonest miners}
and hold them accountable.
In the next \secref{analysis},
we will analyse this game in detail.  It will turn out
in \thref{Pd bound} that its Nash equilibrium
is (under certain conditions) indeed such that the miners will
produce blocks that are close to perfectly random instead of withholding
them.  This holds true even if some miner has a stake in a game on the
blockchain as well, and would benefit from manipulating the outcome
of random events.
Thus, the risk and potential damage of block withholding is greatly reduced
by adding the betting game to the blockchain's rules.

Before continuing, let us clarify the terms we will use in the future
for the different roles that participants on the blockchain network have:

\begin{description}

\item[Miners] produce blocks as described in
\subsecref{mining}.  Thus they are also ultimately in charge of determining
the outcome of random events, which are based on their block hashes.
They may also participate in games on the blockchain, and thus have
preferences for certain outcomes of those events.

\item[Users] of the blockchain network are, in the context of the
following discussion, participants in the betting game described above.
They may bet money against miners, particularly if the block distribution
produced by the miners is not fully random.
They may also participate in games on the blockchain, but this is not
relevant for our discussion below.

\item[Players] are simply participants in a blockchain game and they are
not mining themselves.  The main goal of our proposal is to make sure that
players can expect ``fair'' determination of random events in the games
they play.

\end{description}

We assume that all of these actors are rational, and interested
only in maximising their profit from within the system.
For instance, we assume that no-one is discouraged from an action
damaging the blockchain ecosystem (like manipulating randomness)
just because it may lower the market value of the underlying cryptocurrency
in which they may hold a stake.

\mysubsection{simplification}{Simplifying Assumptions}

To simplify our analysis, let us assume that there is only one miner
(who consequently produces all the blocks).  Similarly, we consider only
one betting user in the system.  Of course, this is far from what the
reality will be.  But if there are multiple agents of each role, then
their combined actions will simply amount to a mixed strategy of the
game (described in more detail below in \subsecref{game}).
For instance, if there is one miner who engages in block withholding
and one who does not, then this is equivalent to a single miner
who only withholds some blocks.  Similarly, if different users on the
blockchain bet different amounts (and perhaps on different outcomes),
then only the ``net bet'' is important for the miners.

Furthermore, the mixed system will tend towards the same equilibrium
as the game with only one miner and one user:  If \emph{overall}, miners
are withholding blocks too often, then it will encourage users (one or multiple)
to bet more against them.  And then either some of the withholding miners
will turn honest (since they are losing money on those bets), or perhaps
more honest miners will join (as they can benefit from their house edge).
In both cases, the overall fraction of withholding will decrease until the
equilibrium is reached, just as a single, rational miner would do.
In the same way, overall betting will converge towards the same equilibrium
that a single, rational betting user would choose.

Hence, by restricting ourselves to just one miner and a single user (who can
and will choose suitable mixed strategies), the overall structure of the
game is not changed.
(A different way to arrive at the same simplifying assumption is the following:
Even if there are multiple miners or users in the system, all of them
will have the same incentives and thus behave in the same way, as long
as we assume all agents to be rational and interested only
in maximising their profits.)

The second simplification we want to make concerns the set of in-game events
we consider as targets of miner manipulation and for users to bet on.
It is likely that any sufficiently complex blockchain game will
need \emph{many different} random events in its state-transition rules.
And while users can then of course bet independently on the different
events as they wish, a miner's decision to publish or withhold a block
is ``atomic'' and cannot be made independently for the different events.
But in the end, all that matters to players is that all events relevant
to them have the distribution they should have.  And if they do not,
then users will have an incentive to bet against miners until this is fixed.
Individually for each event, this game follows our analysis of
\secref{analysis}.

\mysubsection{practical issues}{Considerations for a Practical Implementation}

It is important to note that it is far from straight-forward to
build a practical blockchain that incorporates an implementation
of the betting game described above.  At least the following
issues need to be considered and solved for that:

\begin{itemize}

\item
User bets will have to be submitted to the network as transactions.
They could use some form of hash commitment to hide details about the bet
from miners, but miners may still try to censor those transactions (i.e. simply
not include them in blocks).  Particularly miners who want to manipulate
random events may try to do this and block users from holding them
accountable.  But as long as a user can bet on blocks far enough into
the future, it is enough if \emph{any} (honest)
miner confirms her transaction
before her bet's target block.

\item
Even if honest miners win on average in the betting game (thanks to the
house edge), this scheme still increases variance in mining payout and
risk for the miners.  Thus it may discourage people from mining---although
given that the payout expectation is positive, that remains to be seen.
Miners on such a blockchain are a mixture between miners of a classical
blockchain and casino operators.  Since people currently are providing
both kinds of services, it seems plausible that there will also be miners
on a blockchain based on our proposal.

\item
It is not trivial how users can actually win money from the miners,
especially if the amounts are larger than individual block rewards.
To implement this, it will likely be necessary for miners to deposit a large
stake of coins in order to be able to produce blocks.  Then winnings
of users could be taken out of that deposit.
This changes the structure of the underlying system significantly
(compared to existing PoW blockchains), but does not pose any
problems that are impossible to solve.
It is important to note here that such a blockchain would still
be secured by PoW.  Even though miners are required to stake a deposit,
this would be very different from the existing concept
of proof-of-stake mining.

\item
For generic blockchains like Ethereum or XAYA, there needs to be some
mechanism (e.g. EVM bytecode) to actually define the events that
users bet on.  For games with their own custom blockchain (e.g. Huntercoin),
the developers can instead predefine a list of events
that are likely of interest to players in the game.

\end{itemize}

All in all, defining and building a suitable practical implementation
requires additional research and engineering, but is certainly not impossible.
This, however, is outside the context of
the current paper.  Here, we just want to describe the basic idea and
show that it is---in theory---able to align the game-theoretic
incentives correctly.

\mysection{analysis}{Game-Theoretic Analysis}

Let us now take a detailed look at the game-theoretic incentive structure
that the proposed betting game from \secref{betting} has.

\mysubsection{setting}{Basic Setting}

Let $(\Omega, \F, \P)$ be a probability space.
(For a general introduction to mathematical probability theory, see,
for instance, Chapter~4 of \cite{ashProbability}.)
This probability space models the outcome of mining one block
under the assumption that no block withholding is taking place.
For a blockchain that uses some $b$-bit hash $h$ based on the block to determine
randomness in games, a typical setting will be
\begin{equation*}
\Omega = \setr{h}{h \in \N_0, \; 0 \le h < 2^b}, \;\;
\F = 2^\Omega, \;\;
\prob{A} = \frac{\abs{A}}{2^b}.
\end{equation*}
Here, $2^\Omega$ denotes the power set of $\Omega$ and
$\abs{A}$ is the number of elements in the finite set $A \subset \Omega$.
It is easy to see that this defines, indeed, a (discrete) probability space.
The exact nature of the probability space is not relevant for our further
analysis, though, and can be left unspecified.

Now, let $\D \in \F$ be some fixed event that matters in the blockchain game.
In the following, we will analyse the betting game based on this event.
For instance, this could be the set of outcomes that lead to some particularly
important in-game event in an MMO like Huntercoin, or it could be the set
of ``winning block hashes'' for a casino game like SatoshiDice.
Let us denote the probability of $\D$ by $p \ldef \prob\D$.

Next, we consider the reward $R$ that the miner gets for the block---this
includes his block reward, but it may also include some winnings (or losses)
from in-game events for a miner who also participates in games.
Overall, $R$ is a random variable on our probability space.  Since our analysis
is focused around $\D$, let us define the expected miner rewards related to
the outcome of this event:
\begin{equation}
\label{eq:miner rewards}
R_d \ldef R_0 + R_w \ldef \expect{\cond{R}{\D}}, \;\;
R_n \ldef R_0 \ldef \expect{\cond{R}{\Omega \setminus \D}}
\end{equation}
Here, $R_d, R_n \in \R$ are the expected rewards of the miner for blocks
that trigger $\D$ or not, respectively.  Without loss of generality, we
can assume $R_d \ge R_n$, i.e. that $\D$ is beneficial to the miner.
Since $R_d = R_n$ would make any block withholding and our entire analysis
here pointless, we assume furthermore $R_d > R_n$.
As indicated in \eqref{miner rewards}, we can split the expectation
values into an unconditional base block reward $R_0$
and potential winnings $R_w > 0$ in the game when $\D$ occurs.

\mysubsection{game}{The Betting Game}

The process of mining a new block and potentially betting on the outcome
of $\D$ against the miner (as described in \secref{betting}) can now be seen
as a finite game between two players, the miner and the betting user.

\subsubsection{Choices for the Miner}

For the purpose of this analysis, we assume that before constructing the
next block, the miner decides on one of three possible pure strategies:

\begin{description}

\item[$H$]
The miner can be \emph{honest}, which means that they will simply broadcast
the next block they find, independently of the outcome of $\D$.

\item[$W_d$]
The miner will only broadcast a block where $\D$ occurs.  He will withhold
all blocks that trigger $\Omega \setminus \D$ instead.
(In other words, the miner \emph{withholds to force $\D$}, not
withholds blocks with $\D$.)

\item[$W_n$]
The miner will only broadcast a block where $\D$ does not occur.
In other words, blocks with $\D$ will be withheld and the next published
block will trigger $\Omega \setminus \D$.
Naively, this strategy has no benefit for the miner---but since it is a
valid choice, we nevertheless include it in our analysis.

\end{description}

Since we assume that we only have one miner, they are able to force the outcome
of $\D$ for the next block if they wish (by retrying as often as necessary
to find a suitable block).  Since creating a block likely incurs a cost
to the miner (except perhaps when merge mining), choosing $W_d$ or $W_n$
as a strategy has an extra cost.  This will be reflected in our payoff
function \eqref{miner payoff} below.

Instead of choosing a pure strategy, the miner can of course also pick
a \emph{mixed strategy} that randomises between the three available
pure strategies.  For this case, let us denote the probabilities of the miner
choosing $W_d$ and $W_n$ by $\omega_d$ and $\omega_n$, respectively.
Then the set of possible strategies for the miner is
\begin{equation*}
S_m = \setr{(\omega_d, \omega_n)}
           {\omega_d, \omega_n \in [0, 1], \; \omega_d + \omega_n \le 1}.
\end{equation*}
Clearly, the probability for the miner choosing $H$ is
$1 - \omega_d - \omega_n$.  By definition of $S_m$, this value is also
non-negative.

Based on which strategy from $S_m$ the miner chooses, the distribution
of published blocks may be different from the underlying probability space.
In particular, let us define $P_d$ and $P_n$ to be the probabilities of a mined
block triggering $\D$ and $\Omega \setminus \D$, respectively,
\emph{under the chosen miner strategy from $S_m$}.  It is easy to see
that these quantities are given by
\begin{equation}
\label{eq:Pd}
P_d = \omega_d + (1 - \omega_d - \omega_n) p, \;\;
P_n = \omega_n + (1 - \omega_d - \omega_n) (1 - p) = 1 - P_d.
\end{equation}

\subsubsection{Choices for the Betting User}

The user betting against the miner has also three pure strategies
available:

\begin{description}

\item[$A$]
The user can abstain from betting on the outcome of $\D$ altogether.

\item[$B_d$]
The user bets a maximum amount $b_d > 0$ on $\D$ occurring.

\item[$B_n$]
The user bets a maximum amount $b_n > 0$ on $\Omega \setminus \D$ occurring.
As before, this strategy seems not very useful
at least from a naive point of view (as also the miner
has no incentive to force blocks that do not trigger $\D$), but it is a
possible strategy to consider.

\end{description}

Just like the miner, the betting user can now also employ a mixed strategy.
We denote the probabilities of choosing $B_d$ and $B_n$ by $\lambda_d$
and $\lambda_n$, respectively, then the set of possible user strategies
is given by
\begin{equation}
\label{eq:user strategies}
S_u = \setr{(\lambda_d, \lambda_n)}
           {\lambda_d, \lambda_n \in [0, 1], \; \lambda_d + \lambda_n \le 1}.
\end{equation}
As before, the non-negative probability of choosing $A$ instead is
$1 - \lambda_d - \lambda_n$.

Note that opting for a mixed strategy of, say, betting a fixed amount $b_d$
with probability $\lambda_d$ is (on average)
equivalent to always betting a reduced amount $\lambda_d b_d$.
(This can be seen from \eqref{user payoff} below.)
So instead of defining $S_u$ as in \eqref{user strategies}, we could directly
define the user strategy as a pair of non-negative amounts bet on $\D$
and $\Omega \setminus \D$.  But the definition based on a finite amount of
available pure strategies and mixed strategies based on them fits better to
the typical structure of a game-theoretic analysis.
Hence we decided to stick to this form.

\subsubsection{Payoff Functions}

Let us now take a closer look at the bets that the user can make
against the miner.  As mentioned above, when playing the pure strategy
$B_d$, the user bets $b_d$.  This means that she loses $b_d$ if the bet
fails ($\D$ does not occur).  If, on the other hand, $\D$ takes place,
then she should win an appropriate amount such that the bet is fair
(taking the probability $p$ for $\D$ into account) and includes
a certain house edge $\epsilon > 0$.
First of all, the winning amount should obviously be proportional
to the bet.  In other words, let the winning amount be $\beta_d b_d$ with
some factor of proportionality $\beta_d > 0$.  The correct factor can then be
determined easily, assuming that we want an honest miner to win
on average according to the house edge $\epsilon$:

\begin{lemma}{beta factor}
Consider a simple betting game where a user loses the bet $b > 0$
with probability $1 - p$ and wins $\beta b$ with probability $p > 0$.
If $\beta$ is chosen as
\begin{equation*}
\beta = \frac{1 - p}{p + \epsilon},
\end{equation*}
then the expected win is $\expect{W} = -\epsilon \beta b$.

\begin{proof}
We can simply check that the expectation value is as claimed:
\begin{equation*}
\expect{W}
  = p \beta b - (1 - p) b
  = \frac{(1 - p) p - (1 - p) (p + \epsilon)}{p + \epsilon} \cdot b
  = -\frac{(1 - p) \epsilon}{p + \epsilon} \cdot b
  = -\epsilon \beta b
\end{equation*}
\end{proof}
\end{lemma}

Thus, following \lemmaref{beta factor}, we define
\begin{equation}
\label{eq:beta factors}
\beta_d = \frac{1 - p}{p + \epsilon}, \;\;
\beta_n = \frac{p}{1 - p + \epsilon}.
\end{equation}
This yields the following expression for the expected payoff of the
betting user, based on her strategy $(\lambda_d, \lambda_n) \in S_u$
and the probabilities $P_d$ and $P_n$ that depend on the miner's strategy:
\begin{equation}
\label{eq:user payoff}
\expect{U}
  = \lambda_d b_d \left(\beta_d P_d - P_n\right)
      + \lambda_n b_n \left(\beta_n P_n - P_d\right)
\end{equation}

For the miner, let us assume that the expected cost of ``forcing''
a block that triggers $\D$ by withholding all other blocks is given
by $C_d \ge 0$.  Similarly, $C_n \ge 0$ shall be the cost of forcing
a block where $\D$ does not occur.
The values of these parameters depend on $p$ and the concrete situation of
the blockchain, e.g. the mining difficulty and whether or not the
blockchain is merge mined.
Overall, the payoff for the miner consists of three parts:
First, the expected rewards $R_d$ and $R_n$ for blocks where
$\D$ and $\Omega \setminus \D$ occur, respectively.  Second, the
betting game with the user---since this is a zero-sum game, the miner's
payoff is exactly $-U$.  And third, the costs for forcing a certain
outcome through block withholding.
Taking all together, the expected payoff for the miner is
\begin{equation}
\label{eq:miner payoff}
\expect{M}
  = P_d R_d + P_n R_n - \expect{U} - \omega_d C_d - \omega_n C_n.
\end{equation}

\mysubsection{user strategy}{Strategy for the Betting User}

Since we assume that the betting user acts rationally and purely based on
maximising her profit, let us now consider what strategy choice in $S_u$
maximises $\expect{U}$.  For this, we assume that some miner strategy from
$S_m$ is fixed, and that the resulting probabilities $P_d$ and $P_n = 1 - P_d$
are known to the user.  In a real-world setting, these probabilities can
simply be determined empirically from an analysis of the last blocks.
A first, trivial conclusion is that there exists a (not necessarily unique)
optimal user strategy for any given miner strategy
(we will fully characterise
it later in \corref{optimal user strategy}):

\begin{lemma}{user optimum exists}
For any $P_d \in [0, 1]$, there exists a user strategy
$(\lambda_d^*, \lambda_n^*) \in S_u$ that maximises $\expect{U}$
over $S_u$.

\begin{proof}
This is immediately clear, since $S_u$ is compact and $\expect{U}$ continuous.
(See, for instance, Theorem~2.10 in \cite{rudinAnalysis}.)
\end{proof}
\end{lemma}


From \eqref{user payoff}, it is easy to see that the best choice of
$(\lambda_d, \lambda_n) \in S_u$ depends directly on the signs of the
two terms $\beta_d P_d - P_n$ and $\beta_n P_n - P_d$.  When any one of them
is negative, then the corresponding $\lambda$ should be chosen as zero to
maximise $\expect{U}$.  Otherwise, it should be chosen as large as possible.
These terms are visualised in \figref{user payoff diffs}.  If the miner-chosen
$P_d$ is close to the ``natural value'' of $p = \num{0.4}$,
then both terms are negative---this reflects
the house edge.  But if $P_d$ diverges sufficiently from $p$ to either side,
then the user's profits from betting on this divergence exceed the house edge.
Clearly, the divergence required to make the terms positive is bigger
for a larger house edge (as in \subfigref{user payoff diffs}{large eps}).
In the limit $\epsilon \to 0^+$, the lines intersect at $(p, 0)$.
Let us also give a more formal and detailed analysis:

\includesubfig{user payoff diffs}
  {The two terms $\beta_d P_d - P_n$ (blue) and $\beta_n P_n - P_d$ (red)
   for $p = \num{0.4}$.  The probability $P_d$, which reflects the
   miner's strategy, is on the $x$-axis.}
  {
    \subfig{0.45}{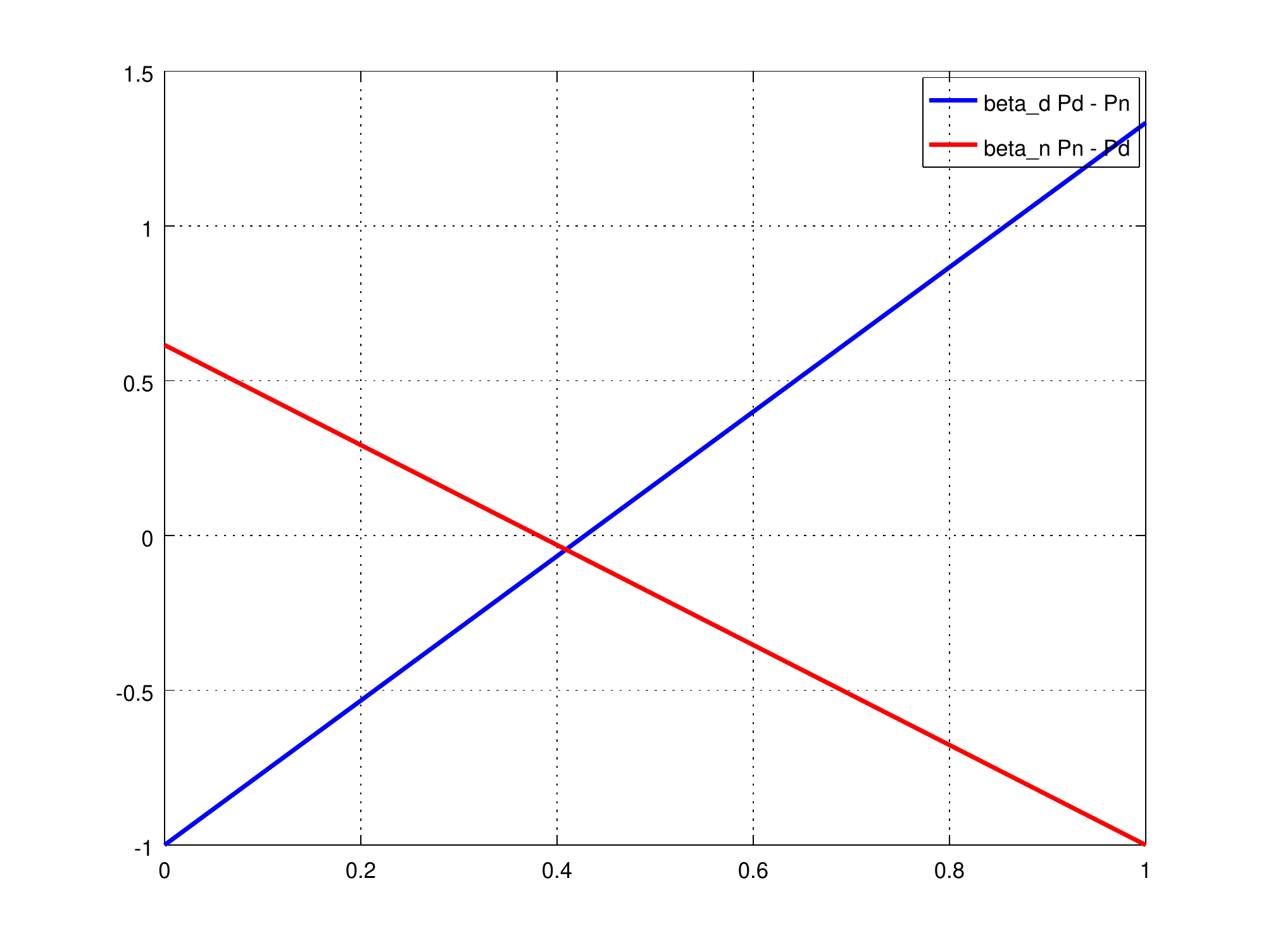}{small eps}
        {Small house edge, $\epsilon = \num{0.05}$}
    \subfig{0.45}{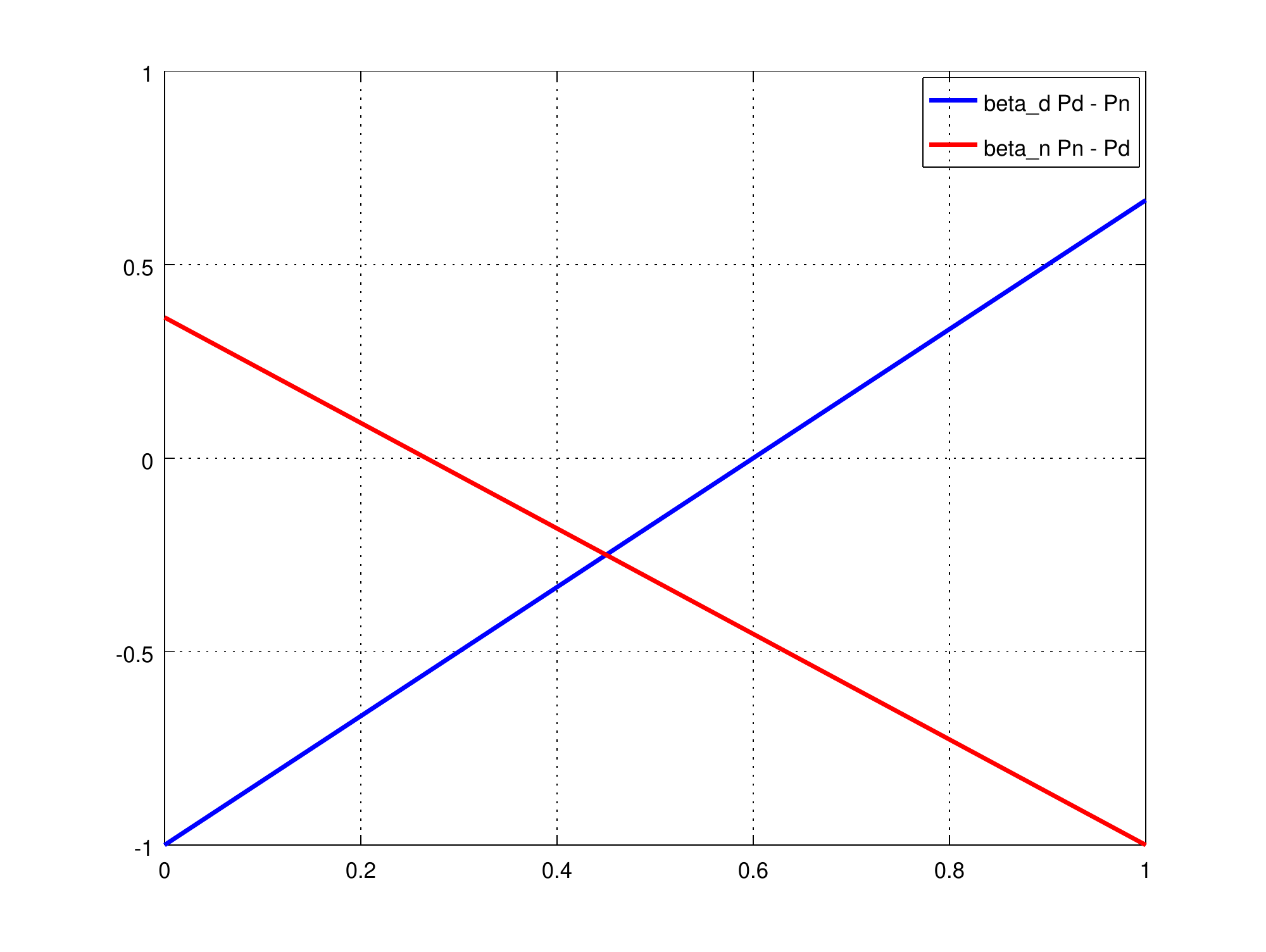}{large eps}
        {Large house edge, $\epsilon = \num{0.5}$}
  }

\begin{lemma}{user payoff difference terms}
Let $\epsilon > 0$, $p \in (0, 1)$ and $P_d \in [0, 1]$.  We set
\begin{equation}
\label{eq:user payoff difference terms Pds}
\underline{P_d} = \frac{\beta_n}{1 + \beta_n} \text{ and }
\overline{P_d} = \frac1{1 + \beta_d},
\end{equation}
where $\beta_d$ and $\beta_n$ are defined as in \eqref{beta factors}.
Then
\begin{equation}
\label{eq:user payoff difference terms ineq}
0 < \underline{P_d} < p < \overline{P_d} < 1.
\end{equation}
In the limit for vanishing or very large house edge, some of these
inequalities turn into equations:
\begin{equation*}
\lim_{\epsilon \to 0^+} \underline{P_d}
  = \lim_{\epsilon \to 0^+} \overline{P_d}
  = p, \;\;
\lim_{\epsilon \to \infty} \underline{P_d} = 0, \;\;
\lim_{\epsilon \to \infty} \overline{P_d} = 1
\end{equation*}

Furthermore,
\begin{equation}
\label{eq:user payoff difference terms signs}
\sign{\beta_d P_d - P_n} = P_d - \overline{P_d}, \;\;
\sign{\beta_n P_n - P_d} = \underline{P_d} - P_d.
\end{equation}
Here, $\sign\cdot$ denotes the three-valued sign function (with
$\sign0 = 0$).

\begin{proof}
Combining \eqref{beta factors} with \eqref{user payoff difference terms Pds},
we find that
\begin{equation*}
\underline{P_d}
  = \frac{p}{1 - p + \epsilon} \cdot \frac{1 - p + \epsilon}{1 + \epsilon}
  = \frac{p}{1 + \epsilon}, \;\;
\overline{P_d}
  = \left(\frac{1 - p + p + \epsilon}{p + \epsilon}\right)^{-1}
  = \frac{p + \epsilon}{1 + \epsilon}.
\end{equation*}
From these representations,
\eqref{user payoff difference terms ineq} and the stated limits
follow easily.

It remains to show \eqref{user payoff difference terms signs}.
For this, note that the following inequalities are equivalent:
\begin{equation*}
\beta_d P_d - P_n \ge 0
  \;\; \Leftrightarrow \;\;
\beta_d P_d \ge 1 - P_d
  \;\; \Leftrightarrow \;\;
P_d \ge \frac1{1 + \beta_d}
  \;\; \Leftrightarrow \;\;
P_d - \overline{P_d} \ge 0
\end{equation*}
Since the same is true for strict inequalities, the first relation
in \eqref{user payoff difference terms signs} follows.
Similarly,
\begin{equation*}
\beta_n P_n - P_d \ge 0
  \;\; \Leftrightarrow \;\;
\beta_n (1 - P_d) \ge P_d
  \;\; \Leftrightarrow \;\;
\frac{\beta_n}{1 + \beta_n} \ge P_d
  \;\; \Leftrightarrow \;\;
\underline{P_d} - P_d \ge 0.
\end{equation*}
Again, this also holds true for strict inequalities,
completing the proof.
\end{proof}
\end{lemma}


The values $\underline{P_d}$ and $\overline{P_d}$ from
\lemmaref{user payoff difference terms}
correspond to the points where the red and blue
lines in \figref{user payoff diffs}, respectively, intersect the
$x$-axis.  They determine the optimal user strategy:

\begin{corollary}{optimal user strategy}
The set $S_u^* \subset S_u$ of user strategies that maximise
$\expect{U}$ is given as follows:

\begin{itemize}

\item
If $P_d < \underline{P_d}$, then the user can make a profit by betting
on $\Omega \setminus \D$.  Consequently, the unique optimal strategy is
to choose $\lambda_d = 0$ and $\lambda_n = 1$,
i.e. $S_u^* = \directset{(0, 1)}$.

\item
If $\underline{P_d} < P_d < \overline{P_d}$, then the user should not bet
at all.  In this range, the distribution of blocks produced by the miner is
so close to the ``true'' distribution (determined by $p$) that the house edge
is larger than any potential winnings.
In this case, $S_u^* = \directset{(0, 0)}$.

\item
If $\overline{P_d} < P_d$, then the user can make a profit by betting
on $\D$.  The user should choose $\lambda_d = 1$ and $\lambda_n = 0$
to maximise $\expect{U}$, i.e.
$S_u^* = \directset{(1, 0)}$.

\item
For the two ``intermediate'' cases $P_d = \underline{P_d}$ and
$P_d = \overline{P_d}$, the user can choose to bet or not, since the
expected winnings from the skewed distribution will exactly compensate
for the house edge.
Thus all strategies with $\lambda_d = 0$ and
$\lambda_n \in [0, 1]$ are optimal for $P_d = \underline{P_d}$,
i.e. $S_u^* = \directset{0} \times [0, 1]$.
Similarly, $S_u^* = [0, 1] \times \directset{0}$ for $P_d = \overline{P_d}$.

\end{itemize}

\begin{proof}
This follows from \eqref{user payoff} and
\lemmaref{user payoff difference terms}.
\end{proof}
\end{corollary}

Note that \eqref{user payoff difference terms ineq} implies
that it is never optimal for the user to bet on both outcomes at the
same time (i.e. $\lambda_d > 0$ and $\lambda_n > 0$).  ``Hedging the bet''
like this always means that she loses unnecessarily much to the house edge.

\mysubsection{miner strategy}{Strategy for the Miner}

Also the miner is trying to optimise his profit, i.e. to maximise
$\expect{M}$ by choosing $(\omega_d, \omega_n) \in S_m$ in the right way
for a given user strategy in $S_u$.
Noting that $P_n = 1 - P_d$ and using $R_0$ and $R_w$ to express the
block rewards, we can rewrite \eqref{miner payoff} to:
\begin{equation}
\label{eq:miner payoff Rw}
\expect{M}
  = P_d (R_0 + R_w) + (1 - P_d) R_0 - \expect{U} - \omega_d C_d - \omega_n C_n
  = R_0 + P_d R_w - \expect{U} - \omega_d C_d - \omega_n C_n
\end{equation}
Since $R_0$ is just a constant offset, it is clear that we can, without
loss of generality, assume $R_0 = 0$ and use the simplified form
\eqref{miner payoff Rw} for analysing the maximum.
(The base block reward $R_0$ is of course important to incentivise miners
in the first place.  But for our analysis here, we assume that there is
a miner on the blockchain anyway.  For determining his strategy with respect
to block withholding, the value of $R_0$ does not matter.)


A first observation we can make about \eqref{miner payoff Rw} is the
following:  The first part of the expression depends only on
$P_d$, but not directly on $\omega_d$ or $\omega_n$.  Only the last two
terms (expressing the cost for withholding blocks) are given based on
$\omega_d$ and $\omega_n$ directly, and they are always ``bad'' for
maximising the profit.  Furthermore, the miner can achieve an
arbitrary value of $P_d \in [0, 1]$ with at least one of the two
withholding probabilities set to zero:

\begin{lemma}{Pd with one omega zero}
Let a desired $P_d \in [0, 1]$ be given arbitrarily.
This value of $P_d$ can be achieved according to \eqref{Pd} with
a miner strategy $\omega^* = (\omega_d^*, \omega_n^*) \in S_m$, where
$\omega_d^* = 0$ if $P_d \le p$ and $\omega_n^* = 0$ if $P_d \ge p$.

Furthermore, any other strategy $(\omega_d, \omega_n) \in S_m$ that
yields $P_d$ satisfies $\omega_d > \omega_d^*$ and $\omega_n > \omega_n^*$.

\begin{proof}
Setting $\omega_d^* = 0$ and solving \eqref{Pd} for $\omega_n^*$
yields $\omega_n^* = 1 - P_d / p$.  For $P_d \le p$, we have
$\omega_n^* \in [0, 1]$, so that $\omega^* \in S_m$.
Similarly, $\omega_n^* = 0$ and $\omega_d^* = (P_d - p) / (1 - p)$ also
yield a valid strategy if $P_d \ge p$.
In the special situation of $P_d = p$, both cases result in
$\omega_d^* = \omega_n^* = 0$.

Now let $\omega = (\omega_d, \omega_n) \in S_m$ be another strategy that results
in the given value of $P_d$.  For $P_d \le p$, clearly
$\omega_d \ge \omega_d^* = 0$.  Also, solving \eqref{Pd} for $\omega_n$,
we get
\begin{equation*}
\omega_n = 1 - \frac{P_d}p + \frac{1 - p}p \cdot \omega_d
  = \omega_n^* + \frac{1 - p}p \cdot \omega_d \ge \omega_n^*.
\end{equation*}
If $\omega_d = \omega_d^* = 0$ would be the case, then also
$\omega_n = \omega_n^*$ follows.  This contradicts the assumption that
$\omega^* \ne \omega$.  Thus $\omega_d > \omega_d^*$ and
$\omega_n > \omega_n^*$ must in fact be true.

For $P_d \ge p$, a similar argument shows $\omega_n > \omega_n^* = 0$ and
\begin{equation*}
\omega_d = \omega_d^* + \frac{p}{1 - p} \cdot \omega_n > \omega_d^*.
\end{equation*}
\end{proof}
\end{lemma}

Thus, we can conclude that the optimal miner strategy will be found
in the set
\begin{equation*}
S_m^0 = \left([0, 1] \times \directset0\right)
          \cup \left(\directset{0} \times [0, 1]\right)
      \subset S_m,
\end{equation*}
where at least one withholding probability is zero:

\begin{corollary}{miner strategy one zero}
$\expect{M}$ has a maximum over the set $S_m$.  This maximum can be achieved
even with $\omega^* \in S_m^0$.
If $C_d, C_n > 0$ are strictly positive, then only strategies from
$S_m^0$ can maximise $\expect{M}$.

\begin{proof}
Since $S_m$ is compact and $\expect{M}$ continuous, it is clear that there
exists a maximising strategy $\omega \in S_m$.  This strategy has a
corresponding optimal value $P_d(\omega) \in [0, 1]$.
Thus, \lemmaref{Pd with one omega zero} implies that there exists
$\omega^* \in S_m^0$ with $P_d(\omega^*) = P_d(\omega)$.
And since $\omega \ge \omega^*$ component-wise according to the lemma,
it follows from $C_d, C_n \ge 0$ and \eqref{miner payoff Rw}
that $\expect{M(\omega)} \le \expect{M(\omega^*)}$.
If $C_d, C_n > 0$, then $\omega = \omega^*$ must be the case,
since otherwise $\expect{M(\omega)} < \expect{M(\omega^*)}$ and that
contradicts our assumption of $\omega$ being a maximum.
\end{proof}
\end{corollary}


Note that $\expect{M}$ is an affine function with respect to
$\omega_d$ and $\omega_n$.  The coefficients depend on the user
strategy $(\lambda_d, \lambda_n) \in S_u$ and the parameters like
$R_w$ or $C_d$.  Thus, one can in theory easily compute where exactly on
$S_m^0$ the optimal miner strategy lies once those coefficients
are fixed.  In full generality, however, this is a bit messy and not very
enlightening.  In the following we will instead consider special cases
that yield more qualitative conclusions as well as results relevant for
determining the Nash equilibrium of the betting game later in
\subsecref{equilibrium}.

The first property that we can deduce is quite intuitive:  If the user bets on
$\Omega \setminus \D$, then the miner should certainly not try to force
this outcome.  Not only will he lose money to the user, he will also
have costs for doing so and result in lower block rewards than not
withholding any blocks (or forcing $\D$).

\begin{lemma}{miner should not force n if user bets on n}
Let the user strategy satisfy $\lambda_d = 0$
and assume that $\omega^* = (\omega_d^*, \omega_n^*) \in S_m^0$
maximises $\expect{M}$.
Then necessarily $\omega_n^* = 0$.

\begin{proof}
Assume to the contrary that $\omega_n^* > 0$.  Since $\omega^* \in S_m^0$,
this implies $\omega_d^* = 0$.  Hence, $P_d(\omega^*) < p = P_d(\omega_0)$,
where $\omega_0 = (0, 0) \in S_m^0$ corresponds to the miner strategy
$H$ of being fully honest and not withholding any blocks.

For $\lambda_d = 0$ and $\omega_d = 0$,
the miner payoff \eqref{miner payoff Rw} simplifies to
\begin{equation*}
\expect{M}
  = P_d R_w - \expect{U} - \omega_n C_n
  = P_d R_w - \lambda_n b_n (\beta_n P_n - P_d) - \omega_n C_n.
\end{equation*}
The first two terms are strictly increasing in $P_d$ and the last is
decreasing in $\omega_n$.
Hence, it follows that $\expect{M(\omega^*)} < \expect{M(\omega_0)}$,
contradicting optimality of $\omega^*$.
\end{proof}
\end{lemma}

Of course, since $\D$ is beneficial to the miner, one may expect
the miner to produce blocks that trigger $\D$ more often rather than less
often, i.e. $P_d \ge p$.  Hence, the natural strategy for the betting user
(according \corref{optimal user strategy})
likely has $\lambda_n = 0$ rather than $\lambda_d = 0$.
In this case, the optimal miner strategy depends on the exact relation
between the benefit $R_w$ that $\D$ has for the miner,
the cost $C_d$ for forcing $\D$ and the user's bet $\lambda_d$.
In particular:

\begin{proposition}{miner strategy for d bets}
Let $\lambda_n = 0$, $C_d, C_n > 0$ and define
\begin{equation}
\label{eq:miner strategy for d bets defs}
\Lambda \ldef \lambda_d b_d (\beta_d + 1), \;\;
\underline\Lambda \ldef R_w - \frac{C_d}{1 - p}
  < R_w + \frac{C_n}p \rdef \overline\Lambda.
\end{equation}
Then the set $S_m^* \subset S_m$ of miner strategies that maximise $\expect{M}$
is given as follows:

\begin{itemize}

\item
If $\Lambda < \underline\Lambda$, then the rewards of blocks with $\D$
outweigh the costs of forcing them.  The miner should choose $\omega_d = 1$
and $\omega_n = 0$, i.e. $S_m^* = \directset{(1, 0)}$.

\item
If $\underline\Lambda < \Lambda < \overline\Lambda$, then the user bets
roughly equal the rewards $R_w$ that the miner can achieve with blocks that
trigger $\D$.  The costs for forcing either $\D$ or $\Omega \setminus \D$
outweigh any benefits, so the optimal miner strategy is to be honest with
$S_m^* = \directset{(0, 0)}$.

\item
If $\overline\Lambda < \Lambda$, then user bets on $\D$ are so high that
the miner can actually benefit the most by forcing blocks that trigger
$\Omega \setminus \D$.  He should choose $\omega_d = 0$ and $\omega_n = 1$,
i.e. $S_m^* = \directset{(0, 1)}$.

\item
For the intermediate cases with equality, the miner can choose any mixed
strategy between the two equal choices.  In other words, for
$\Lambda = \underline\Lambda$, $S_m^* = [0, 1] \times \directset0$.
For $\Lambda = \overline\Lambda$, the optimal strategies are given by
$S_m^* = \directset0 \times [0, 1]$.

\end{itemize}

\begin{proof}
For $\lambda_n = 0$, we can rewrite $\expect{M}$ from \eqref{miner payoff Rw}
as
\begin{equation*}
\expect{M} = X_d \omega_d + X_n \omega_n + X_0,
\end{equation*}
where the coefficients are given by
\begin{equation*}
\begin{split}
X_d & = R_w (1 - p) - C_d - \Lambda (1 - p), \\
X_n & = \Lambda p - R_w p - C_n, \\
X_0 & = R_w p + \lambda_d b_d (1 - p (\beta_d + 1)).
\end{split}
\end{equation*}
From these values, it is easy to see that
$\sign{X_d} = \sign{\underline\Lambda - \Lambda}$
and $\sign{X_n} = \sign{\Lambda - \overline\Lambda}$.
Since $\underline\Lambda < \overline\Lambda$, at most one of the two can
be positive (this matches the result of \corref{miner strategy one zero}).
The optimal miner strategies $S_m^*$ as stated follow easily.
\end{proof}
\end{proposition}

Let us conclude this subsection with two remarks about
\propref{miner strategy for d bets}:
First, note that \eqref{beta factors} implies
\begin{equation*}
X_0 = R_w p + \epsilon \beta_d \cdot \lambda_d b_d > 0.
\end{equation*}
Thus, the miner payoff for being honest is always positive
and corresponds to what one expects from
\lemmaref{beta factor}.
Second, it can of course be the case that not all of the options
in \propref{miner strategy for d bets} are actually applicable.
For instance, if $R_w$ is very small compared to the cost $C_d$, then
$\underline\Lambda < 0$ can be the case.  This means that $W_d$ (forcing
blocks that trigger $\D$) is never a good choice for the miner, even
if $\lambda_d = 0$.

\mysubsection{equilibrium}{Nash Equilibria of the Betting Game}

After having analysed the optimal strategies for both the miner
and betting user previously, we can now consider them together.
To analyse the general structure of the betting game, we employ the
widely-used concept of \emph{Nash equilibria} \cite{nash}.
For a more extensive introduction to this
concept, see Chapter~2 of \cite{osborneRubinstein}.
Roughly speaking, a Nash equilibrium occurs in a game if every player
uses the optimal strategy assuming that all other players stick to their
chosen strategy.  In other words, when all players choose a strategy from
such an equilibrium point, then none of them has an incentive to
unilaterally switch to a different strategy.
In the context of our the betting game, this means
(compare Definition~14.1 in \cite{osborneRubinstein}):

\begin{definition}{Nash equilibrium}
A pair of strategies $(\omega, \lambda) \in S_m \times S_u$ is a
\emph{Nash equilibrium} if $\omega$ maximises $\expect{M(\lambda)}$
over $S_m$ and $\lambda$ maximises $\expect{U(\omega)}$ over $S_u$.
\end{definition}

It is not hard to see that Proposition~20.3 of \cite{osborneRubinstein}
applies in our situation, showing that a Nash equilibrium exists.
In the remainder of this subsection, however, we will characterise the
Nash equilibria of our betting game more thoroughly.
For this, we will assume $C_d, C_n > 0$.  This acts as a kind of
regularisation (see \corref{miner strategy one zero}), and will reduce
the number of special cases we have to consider.


The first result that we can derive is that the ``negative'' strategies
$W_n$ and $B_n$, as well as mixed strategies involving them, can never
be part of a Nash equilibrium.  This already simplifies our analysis
of the equilibria quite a lot.

\begin{lemma}{negative strategies no equilibrium}
Let $(\omega, \lambda) \in S_m \times S_u$ be a Nash equilibrium.
Then necessarily $\omega_n = 0$ and $\lambda_n = 0$.

\begin{proof}
Assume to the contrary that $\omega_n > 0$.  According to
\corref{miner strategy one zero}, this implies $\omega_d = 0$.  Hence,
$P_d < p$ must be the case (i.e. the miner forces $\Omega \setminus \D$
to occur more often than it would naturally).
In this situation, \corref{optimal user strategy} implies
$\lambda_d = 0$ for an optimal user strategy.
But then \lemmaref{miner should not force n if user bets on n}
implies $\omega_n = 0$, which is a contradiction.

Now assume $\lambda_n > 0$ instead.  Since $\lambda$ is a maximising
strategy for the user, this means per \corref{optimal user strategy} that
$P_d < p$ must be the case.  This, in turn, is only possible if $\omega_n > 0$.
But then we arrive at a contradiction as in the first part of the proof.
\end{proof}
\end{lemma}


As our next step, we consider a first special case:  If the reward $R_w$ for
blocks that trigger $\D$ is so small that the cost $C_d$ for forcing them
outweighs any benefits for the miner, then the equilibrium point of the
betting game is with the miner acting honestly and the user not betting at all.
In this case, random numbers will be fair simply because the miner has
no sufficient incentive to fiddle with them.  (Just as they would be fair
in this situation without a betting game.)

\begin{proposition}{equilibrium small Rw}
Assume that $R_w < C_d / (1 - p)$.  Then the unique Nash equilibrium of the
betting game is $\omega_d = \omega_n = \lambda_d = \lambda_n = 0$, i.e.
the pure strategy $(H, A)$.

\begin{proof}
Note that our assumption together with
\eqref{miner strategy for d bets defs} implies $\underline\Lambda < 0$,
so that $\underline\Lambda < \Lambda$ is always the case for
\propref{miner strategy for d bets}.
Let $(\omega, \lambda) \in S_m \times S_u$ be a Nash equilibrium.
Then \lemmaref{negative strategies no equilibrium} implies
$\omega_n = \lambda_n = 0$.
Thus we can conclude from \propref{miner strategy for d bets} that
$\omega_d = 0$ must be the case as well.  Consequently $P_d = p$,
so that $\lambda_d = 0$ follows from \corref{optimal user strategy}.

It remains to verify that $(H, A)$ actually is a Nash equilibrium.
(This follows also from the known existence of an equilibrium, but it is
not hard to check directly.)
For $\omega_d = \omega_n = 0$, $P_d = p$.  Thus $\lambda = (0, 0)$ is
indeed an optimal strategy for the user according to
\corref{optimal user strategy}.
The other way round, for $\lambda_d = \lambda_n = 0$ and
$\underline\Lambda < \Lambda$, the case
$\underline\Lambda < \Lambda = 0 < \overline\Lambda$ is active in
\propref{miner strategy for d bets}, implying that $\omega = (0, 0)$ is
the optimal miner strategy.  This completes the proof.
\end{proof}
\end{proposition}


The second special case is where $R_w$ is large enough to incentivise
block withholding, but where also the maximum bet $b_d$ is so small that
even if the user bets that maximum, then the miner still benefits by
forcing all blocks to trigger $\D$.  In the extreme case $b_d = 0$, this
corresponds to the typical situation where block hashes are used
for random numbers but no betting game is there at all.

\begin{proposition}{equilibrium small bd}
Assume $0 < b_d (\beta_d + 1) < R_w - C_d / (1 - p)$.
Then the unique Nash equilibrium of the betting game is
$\omega_n = \lambda_n = 0$ and $\omega_d = \lambda_d = 1$,
i.e. the pure strategy $(W_d, B_d)$.

\begin{proof}
As before, let $(\omega, \lambda) \in S_m \times S_u$ be a Nash equilibrium
and recall that $\omega_n = \lambda_n = 0$ must be the case according to
\lemmaref{negative strategies no equilibrium}.
In the situation we consider, $\Lambda < \underline\Lambda$ is always the
case (for all possible $\lambda_d \in [0, 1]$).
Thus \propref{miner strategy for d bets} implies $\omega_d = 1$.
Then $P_d = 1$, so that the optimal user strategy is $\lambda_d = 1$
according to \corref{optimal user strategy}.
Using the same results, we can also easily verify that the pair of
pure strategies $(W_d, B_d)$ actually is a Nash equilibrium.
\end{proof}
\end{proposition}


We can now also consider the case between the two previous
extremes.  In that situation, the equilibrium is given by a mixed
strategy:

\begin{proposition}{equilibrium mixed}
Let $0 < R_w - C_d / (1 - p) < b_d (\beta_d + 1)$ be the case.
Then the betting game has a unique Nash equilibrium at a mixed strategy
between $(H, A)$ and $(W_d, B_d)$.  In particular, $\omega_n = \lambda_n = 0$
and
\begin{equation}
\label{eq:equilibrium mixed strategies}
\omega_d = \frac\epsilon{1 + \epsilon}, \;\;
\lambda_d = \frac1{b_d (\beta_d + 1)} \left(R_w - \frac{C_d}{1 - p}\right).
\end{equation}

\begin{proof}
Assume that $(\omega, \lambda) \in S_m \times S_u$ is a Nash equilibrium.
Then \lemmaref{negative strategies no equilibrium} implies
$\omega_n = \lambda_n = 0$.  Consequently also $P_d \ge p$.
Next, consider \eqref{user payoff difference terms Pds}
and note that $P_d = \overline{P_d}$
if and only if $\omega_d$ is chosen as in \eqref{equilibrium mixed strategies}.
Assume for a moment that $\omega_d$ is larger,
which means $P_d > \overline{P_d}$.
Then \corref{optimal user strategy} implies that $\lambda_d = 1$ must be
the case for an equilibrium point as per \defref{Nash equilibrium}.
Hence $\underline\Lambda < \Lambda$ is the case
in \eqref{miner strategy for d bets defs} for the situation we consider.
But then $\omega_d = 0$ according to
\propref{miner strategy for d bets},
which contradicts $P_d > \overline{P_d} > p$.
So consider the situation that $\omega_d$ is smaller,
i.e. $P_d < \overline{P_d}$.
Then \corref{optimal user strategy} implies $\lambda_d = 0$,
so that $\Lambda = 0 < \underline\Lambda$
in \propref{miner strategy for d bets}.  Thus $\omega_d = 1$ must be
the case for an optimal miner strategy, but that contradicts
$P_d < \overline{P_d}$ as it would imply $P_d = 1$.
Hence we have shown $P_d = \overline{P_d}$, which means that
$\omega_d$ must be as in \eqref{equilibrium mixed strategies}.

Similarly, note that $\lambda_d$ matches \eqref{equilibrium mixed strategies}
if and only if $\Lambda = \underline\Lambda$ in
\eqref{miner strategy for d bets defs}.
Assume that $\Lambda < \underline\Lambda$ would be the case.  Then
\propref{miner strategy for d bets} implies $\omega_d = 1$.
For $\underline\Lambda < \Lambda$, $\omega_d = 0$ follows.
Both contradict the already shown form of $\omega_d$ according to
\eqref{equilibrium mixed strategies}, though, so that
$\Lambda = \underline\Lambda$ must be the case.
Thus we have shown
that the Nash equilibrium $(\omega, \lambda)$ necessarily has the
form claimed in \eqref{equilibrium mixed strategies}.

It remains to show that \eqref{equilibrium mixed strategies} is also
sufficient for $(\omega, \lambda)$ being a Nash equilibrium.
For this, note first that $\omega_d$ and $\lambda_d$ from
\eqref{equilibrium mixed strategies} are both valid strategy choices
in $(0, 1)$.
As noted already above, they furthermore imply exactly
$P_d = \overline{P_d}$ and $\Lambda = \underline\Lambda$.
Thus, it follows from \corref{optimal user strategy} that
$\lambda_d$ is optimal for $\omega_d$ (in fact, any $\lambda_d \in [0, 1]$
would be optimal).  Similarly, \propref{miner strategy for d bets}
implies that also $\omega_d$ (and in fact any $\omega_d \in [0, 1]$)
is optimal for $\lambda_d$.
Thus, $(\omega, \lambda)$ is a Nash equilibrium according to
\defref{Nash equilibrium}.
\end{proof}
\end{proposition}

The equilibrium point from \eqref{equilibrium mixed strategies}
can be interpreted as follows:
Due to the positive benefit $R_w$ that the miner has for blocks that
trigger $\D$, he has an incentive to slightly prefer those blocks instead
of producing fully random outcomes.
But since the user is able to bet against the miner, she will do so
to make a profit from the knowledge that $\D$ blocks are more frequent
than they should be.
In the end, the miner skews the distribution just so much that the losses
to the betting user equal the benefits from $R_w$.
For the user, on the other hand, the winnings thanks
to the skewed distribution are just equal to the miner's house edge.
In this situation, the miner benefits compared to a situation without
any block withholding and without any bets, and the user enjoys ``free
betting''.  Furthermore, the situation is even beneficial for miners
that are honest and not interested in block withholding in the first place:
They benefit from a non-zero amount of bets being made; the profit they
make due to the house edge is exactly equal to the benefit they could get
from manipulating the randomness and exploiting $R_w$ instead.


Finally, it remains to consider the situations exactly on the border
between the previous three cases.  For them, it will turn out that
there are multiple equilibrium points.  That is because the user strategy
will be on the boundary of its domain ($\lambda_d \in \directset{0, 1}$).
With active constraints, the user's strategy choice is less flexible,
so that multiple miner strategies can be optimal at the same time.

\begin{proposition}{equilibrium equality cases}
Let us denote the set of Nash equilibria of the betting game
by $\Sigma^* \subset S_m \times S_u$.

If $R_w = C_d / (1 - p)$, then
\begin{equation}
\label{eq:equilibrium equality cases first}
\Sigma^*
    = \setr{(\omega_d, 0, \lambda_d, 0)}
           {\lambda_d = 0, \;\; 0 \le \omega_d \le \frac\epsilon{1 + \epsilon}}.
\end{equation}

For $0 < b_d (\beta_d + 1) = R_w - C_d / (1 - p)$, we have
\begin{equation}
\label{eq:equilibrium equality cases second}
\Sigma^*
    = \setr{(\omega_d, 0, \lambda_d, 0)}
           {\lambda_d = 1, \;\; \frac\epsilon{1 + \epsilon} \le \omega_d \le 1}.
\end{equation}

\begin{proof}
Let us first consider the case $R_w = C_d / (1 - p)$, and let
$(\omega, \lambda) \in S_m \times S_u$ be a Nash equilibrium.
Then clearly $\omega_n = \lambda_n = 0$ according to
\lemmaref{negative strategies no equilibrium}.
Furthermore, in this case we have $\underline\Lambda = 0$.
If $\lambda_d > 0$, then also $\Lambda > 0$ as well.
Hence $\underline\Lambda < \Lambda$ implies $\omega_d = 0$
through \propref{miner strategy for d bets}.
But then $P_d = p$ and thus $\lambda_d = 0$
according to \corref{optimal user strategy}.  This is a contradiction,
so that $\lambda_d = 0$ must necessarily be true.
But $\lambda_d = 0$ is only an optimal strategy according to
\corref{optimal user strategy} if $P_d \le \overline{P_d}$.
This, in turn, is equivalent to $\omega_d \le \epsilon / (1 + \epsilon)$.
Hence, $(\omega, \lambda)$ is indeed in the right-hand side
of \eqref{equilibrium equality cases first}.

Next, let $(\omega, \lambda)$ be in the right-hand side of
\eqref{equilibrium equality cases first}.  We have to show that it is actually
a Nash equilibrium.
Since $\lambda_d = 0$, we know that $\Lambda = \underline\Lambda = 0$.
Thus, any value for $\omega_d$ corresponds to an optimal miner strategy
as per \propref{miner strategy for d bets}.
For $\omega_d \le \epsilon / (1 + \epsilon)$, we know that
$P_d \le \overline{P_d}$ as before.  Hence, $\lambda_d = 0$ is
an optimal strategy for the user according to \corref{optimal user strategy}.

The second case to consider is $b_d (\beta_d + 1) = R_w - C_d / (1 - p)$.
Let $(\omega, \lambda)$ be a Nash equilibrium.
If $\lambda_d < 1$, then $\Lambda < \underline\Lambda$.  Hence
$\omega_d = 1$ must be the case for an optimal miner strategy according to
\propref{miner strategy for d bets}.
But then $P_d = 1$, so that $\lambda_d < 1$ cannot be optimal per
\corref{optimal user strategy}.
Hence, $\lambda_d = 1$ must be the case.
This, however, is only optimal if $P_d \ge \overline{P_d}$, which
in turn is equivalent to $\omega_d \ge \epsilon / (1 + \epsilon)$.
Thus, $(\omega, \lambda)$ must be in the right-hand side
of \eqref{equilibrium equality cases second}.

Finally, consider any $(\omega, \lambda)$ in the right-hand side
of \eqref{equilibrium equality cases second}.
From $\lambda_d = 1$, it follows that $\Lambda = \underline\Lambda$.
Hence, any $\omega_d$ yields an optimal miner strategy as per
\propref{miner strategy for d bets}.
Also, the lower bound that we have on $\omega_d$ implies
$P_d \ge \overline{P_d}$, so that $\lambda_d = 1$ is actually optimal
for the user according to \corref{optimal user strategy}.
Thus, $(\omega, \lambda) \in \Sigma^*$.  This completes the proof.
\end{proof}
\end{proposition}


With the previous results, we have now fully characterised the Nash equilibria
of our betting game in various situations.  Note that the miner withholding
$\omega_d$ is always bounded away from one in a Nash equilibrium, except if
$b_d$ is too small.  But since $b_d$ is just a parameter of the system,
it can be chosen large enough to enable a meaningful betting game against
miner withholding.  Hence, we can conclude that the betting game is
indeed efficient at ensuring that the randomness in our blockchain game
is close to perfectly fair:

\begin{theorem}{Pd bound}
Assume $C_d, C_n > 0$ and that $b_d$ is chosen large enough, i.e. such that
\begin{equation}
\label{eq:Pd bound bd}
b_d (\beta_d + 1) > R_w - \frac{C_d}{1 - p}.
\end{equation}
Then for any Nash equilibrium of the betting game,
\begin{equation}
\label{eq:Pd bound result}
0 \le \omega_d \le \frac\epsilon{1 + \epsilon} \text{ and }
p \le P_d \le \frac{p + \epsilon}{1 + \epsilon}.
\end{equation}
In particular, $P_d$ will be arbitrarily close to $p$ if the
house edge $\epsilon$ is chosen small enough.

\begin{proof}
For the case of \eqref{Pd bound bd}, one of
\propref{equilibrium small Rw}, \propref{equilibrium mixed}
or the first part of \propref{equilibrium equality cases} applies.
All of them yield $\omega_n = 0$ and $\omega_d$ as in \eqref{Pd bound result}.
The bound on $P_d$ from \eqref{Pd bound result} follows then immediately
by \eqref{Pd}.
Finally, it is also easy to see that $P_d \to p$ from above
as $\epsilon \to 0^+$.
\end{proof}
\end{theorem}

\includefig{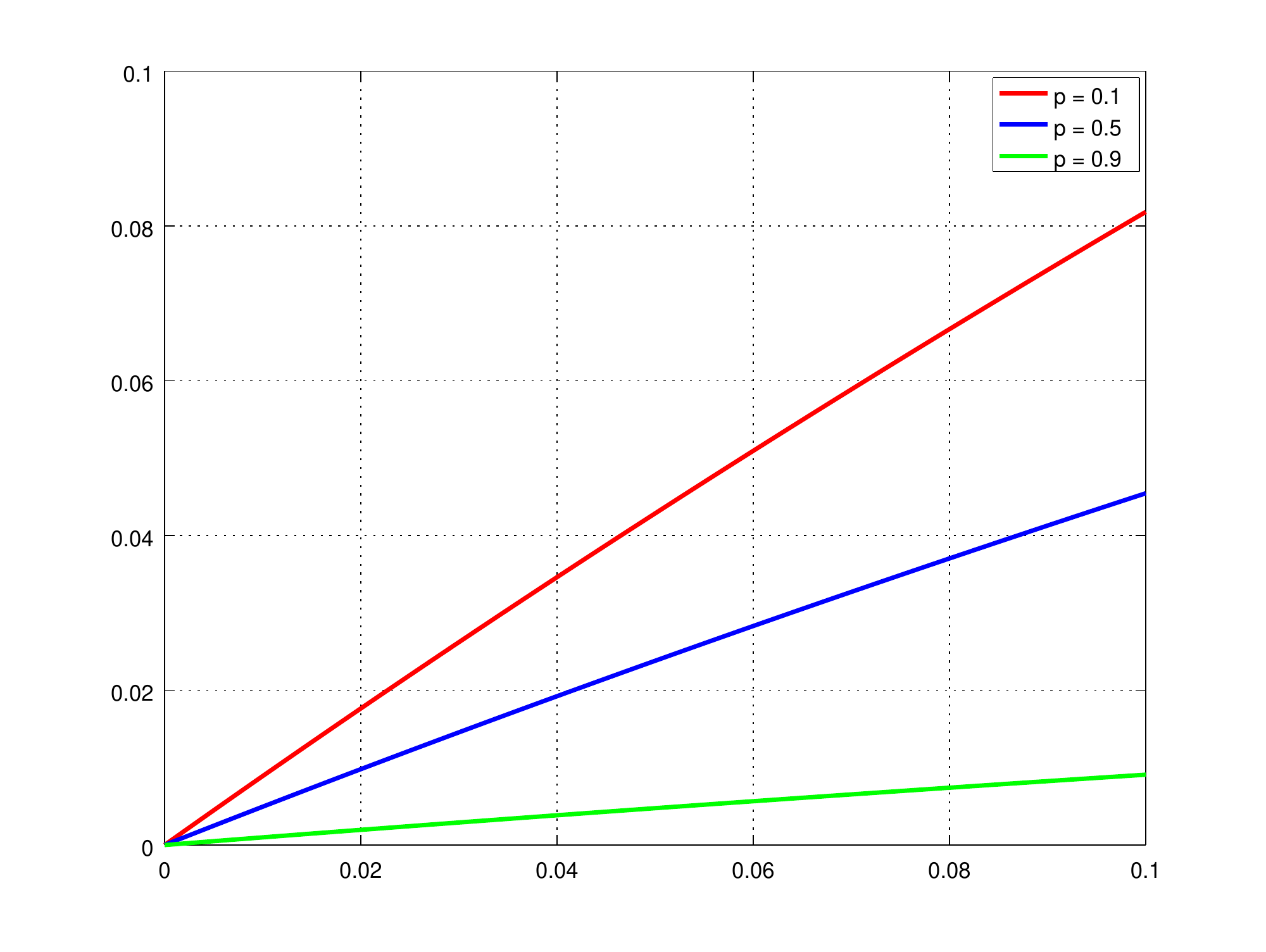}{width=0.6\textwidth}{equilibrium Pd}
  {The deviation of $P_d$ from $p$ in the Nash equilibrium according to
   \thref{Pd bound}.  The house edge $\epsilon$ is on the $x$-axis.
   The $y$-axis shows the upper bound for $P_d - p = \abs{P_d - p}$.}

\figref{equilibrium Pd} shows numerical values for the bounds on
the deviation of $P_d$ from the true $p$ according to \eqref{Pd bound result}.
The three lines correspond to different values of $p$.  It can be seen
that the betting game is most efficient in reducing the miner-created
skew in the distribution of $\D$ if $\D$ is likely to occur, and least
efficient for events that occur rarely.  This can be explained by the fact
that block withholding has the most effect for the miner if the event would
naturally occur only infrequently.
But it can also be seen that even a huge house edge of \percent{10}
only makes a 50-50 event occur \percent{55} instead of \percent{50}
of the time.  For a more competitive house edge of \percent1, this
deviation drops to just slightly more than \percent{0.5}.
For most random events in games, this is barely noticeable---and definitely
a lot better than the situation without betting, where the miner would
(if only interested in profit) force the event to occur \percent{100}
of the time instead.

\mysection{conclusion}{Conclusion}

The current state of the art for randomness in blockchain games relies
on one of two methods, using block hashes to seed PRNGs or basing random
events on hash commitments from the players.  We have discussed that both
approaches have drawbacks under certain circumstances and for certain types
of games.  However, randomness from block hashes can be improved by
introducing a special betting game.  Using game theory, we have shown that
this game leads to Nash equilibria where the observed distribution for
a certain random event matches the expected, fair distribution quite well.
In other words, introducing our betting game changes the incentive structure
for miners in such a way that they will no longer manipulate the randomness
completely.  Instead, the game will be much fairer for every player.

We believe that this is a very interesting result.  It shows that it is
possible to use game-theoretic incentives to produce fair ``randomness'' in
a deterministic context like a blockchain.  Our analysis from \secref{analysis}
has proven the general idea to work.
However, research in this topic is of course not complete yet.
In our opinion, there are two main directions where further work
would be very interesting and useful:

First, we made a lot of assumptions for the analysis here (see
\subsecref{simplification}).  Further work is required to remove those and
check whether our results still hold in more general contexts.
Particularly interesting would be an analysis with more agents than just
one miner and one user.  Also important is work on a situation where there
is more than one event.  For the latter, betting could still be done on
individual events.  But the miner strategy would consist of a full
probability distribution on $\Omega$, likely deviating from the natural
distribution given by $\P$.  A full analysis of this situation will
require different mathematics from the current paper and would be
heavier in topics like functional analysis and measure theory.  Hence, it
should be done in a separate paper.
We believe that this would not change the fundamental results, though.
It is conceivable that there would still be one particular event or outcome
that maximises the miner's benefit $R_w$, so that the miner strategy would
then still be focused on withholding blocks just for this particular event.
Finally, a third assumption we made is that the betting user
actually knows $R_w$.  This is not the case (at least not exactly)
for a real-world implementation, since it depends heavily on how the miner
is involved in the game himself (and if at all).  So the analysis could
be adapted to be more probabilistic in nature, e.g. using Bayesian game theory.
In the end, however, only the value of $P_d$ matters for the user strategy---and
that one can be observed empirically.  So at least if the system still tends
towards some kind of equilibrium, this will yield the same results as
our current analysis.

Second, it remains to implement and test our proposal in a real-world system.
Only that can show whether or not agents will really behave as analysed
and thus produce good randomness.  We have already listed the main issues
to overcome for such an implementation above in \subsecref{practical issues}.
For such a test, it is interesting to note that random numbers can in theory
be separated from mining:  Instead of basing them off a block hash and putting
miners in charge, there could be a separate class of agents that are
just responsible for producing random numbers.  That would, of course, reduce
the costs $C_d$ and $C_n$ drastically and thus make the system even more
susceptible to manipulation.  But the betting game may still be able to
rectify the incentives even in such a system.  The advantage of an
implementation like that is that it could be done on top of an existing
blockchain, e.g. in an Ethereum smart contract or a game on the XAYA platform.
There would be no need to build a blockchain from scratch.
Getting real-world results for how the betting game behaves in such a system
would be very interesting.


\bibliography{references}{}
\bibliographystyle{plain}

\end{document}